\documentclass[twocolumn]{svjour3}  
\smartqed  

\usepackage[dvipdfmx]{color}
\usepackage{graphicx}
\usepackage[T1]{fontenc}
\usepackage{mathptmx}  
\usepackage[scaled]{helvet}  
\usepackage{courier}
\usepackage[subrefformat=parens]{subcaption}

\makeatletter
\let\cl@chapter\undefined
\makeatletter
\usepackage{amsmath,amssymb}   
\usepackage{mathtools} 
\usepackage{cleveref}  
\Crefname{equation}{Eq.}{Eqs.}%
\Crefname{figure}{Fig.}{Figs.}%
\usepackage{amsbsy}
\usepackage{bm}
\usepackage{algorithmic}
\usepackage{algorithm}

\usepackage{enumerate} 

\usepackage{lineno}
\newcommand*\patchAmsMathEnvironmentForLineno[1]{%
  \expandafter\let\csname old#1\expandafter\endcsname\csname #1\endcsname
  \expandafter\let\csname oldend#1\expandafter\endcsname\csname end#1\endcsname
  \renewenvironment{#1}%
     {\linenomath\csname old#1\endcsname}%
     {\csname oldend#1\endcsname\endlinenomath}}%
\newcommand*\patchBothAmsMathEnvironmentsForLineno[1]{%
  \patchAmsMathEnvironmentForLineno{#1}%
  \patchAmsMathEnvironmentForLineno{#1*}}%
\AtBeginDocument{%
\patchBothAmsMathEnvironmentsForLineno{equation}%
\patchBothAmsMathEnvironmentsForLineno{align}%
\patchBothAmsMathEnvironmentsForLineno{flalign}%
\patchBothAmsMathEnvironmentsForLineno{alignat}%
\patchBothAmsMathEnvironmentsForLineno{gather}%
\patchBothAmsMathEnvironmentsForLineno{multline}%
}
\journalname{Journal of Marine Science and Technology}
\begin{document}

\title{Validation of Theoretical Estimation Methods and Maximum Value Distribution Calculation for Parametric Roll Amplitude in Long-Crested Irregular Waves
%
}

\author{Keiji Katsumura         \and Leo Dostal  \and Taiga Kono \\
Yuuki Maruyama         \and Masahiro Sakai \and Atsuo Maki
}

\institute{Keiji Katsumura \and Taiga Kono \and Yuuki Maruyama \and Masahiro Sakai \and Atsuo Maki\at
              Department of Naval Architecture and Ocean Engineering, \\Graduate School of Engineering, Osaka University, \\2-1 Yamadaoka, Suita, Osaka, Japan \\
Leo Dostal \at
Institute of Mechanics and Ocean Engineering, Hamburg University of Technology, 21043 Hamburg, Germany
}

\date{Received: date / Accepted: date}

\maketitle

\begin{abstract}
    Parametric rolling is a parametric excitation phenomenon caused by GM variation in waves. There are a lot of studies of the estimation the conditions, the occurrence, and the amplitude of parametric rolling. On the other hand, there are relatively few cases in which theoretical methods for estimating parametric roll amplitudes in irregular waves have been validated in tank tests. The primary objective of this study is to validate theoretical estimation methods for the parametric roll amplitude in irregular waves and improve their accuracy. First, the probability density functions (PDF) of the parametric roll amplitude obtained from the model ship motion experiment in irregular waves are compared with that obtained from theoretical estimation methods. Second, the method to improve the accuracy of estimation of the roll restoring variation in irregular waves is suggested. Third, the method to estimate the distribution of the maximum amplitude of parametric rolling in irregular waves. As a result, the PDFs of the roll amplitude obtained from the experiments differ from the results of theoretical estimation. After that, by correcting GM variation, the results of theoretical estimation are closer to the experimental results. Moreover, by the theoretical estimation method using the moment equation, the qualitative estimation for the PDF of the maximum roll amplitude is succeeded.

\end{abstract}

\section{Introduction}\label{sec:intro}
    Parametric rolling is a parametric excitation phenomenon caused by GM variation in waves. It is particularly likely to occur on container ships, whose transom stern and bow flare cause significant changes in the secondary moment of the waterline near the stern of the ship. For example, the accident of a C11 class container ship caused by parametric rolling in 1998 is well known~\cite{France2003}. Moreover, the accident of a pure car and truck carrier (PCTC) was also occurred in recent years~\cite{Rosen2012experience}.
    In order to prevent such dangerous phenomenon, parametric rolling, it is necessary to estimate its conditions, occurrence, and amplitude.
    %
\subsection{Related Research}
    There is a long research history on theoretical estimation methods for the conditions, the occurrence, and the amplitude of parametric rolling. The parametric rolling in regular seas has been theoretically explored by Kerwin~\cite{Kerwin1955}, Zavodney et al.~\cite{Zavodney1989}, Francescutto~\cite{Francescutto2001}, Bulian~\cite{Bulian2004approximate}, Spyrou~\cite{Spyrou2005paramet}, Umeda et al.~\cite{Umeda2004nonlinear}, Maki et al.~\cite{Maki2011parametric}, and Sakai et al~\cite{Sakai2018}. Furthermore, in particular, since the 1980s, there have been a lot of studies on theoretical estimation methods for parametric rolling in irregular waves. In order to study parametric rolling in irregular waves, systems excited by colored noise must be treated. One of methods to treat such systems theoretically is through a probabilistic approach. Spyrou~\cite{Spyrou2005paramet}, Dostal~\cite{Dostal2012non} and Maki~\cite{Maki2023} discussed the occurrence of parametric rolling in terms of the destabilization of the origin due to the roll restoring variation in irregular waves. For the estimation of parametric roll amplitudes, studies using stochastic averaging methods have been conducted.  Using the stochastic averaging method, Roberts~\cite{Roberts1982} derived the stochastic differential equation 
    for a phase and an amplitude. In Roberts' results, the probability density function (hereinafter referred to as PDF) of the roll amplitude reflected the characteristics of the damping component and those of the cubic restoring component.  Subsequently, Roberts et al.~\cite{Roberts1986,Roberts2000} proposed an energy based methodology and attempted to reflect the restoring component. Dostal et al.~\cite{Dostal2011} proposed an energy-based stochastic averaging method using the Hamiltonian. Furthermore, Maruyama et al.~\cite{Maruyama2022} developed Dostal's method. They compared the results of Roberts~\cite{Roberts1982} and Dostal~\cite{Dostal2011}'s methods with those of Monte Carlo Simulation (hereinafter referred to as MCS). As a result, they reported a discrepancy in the tail of the PDF and proposed a method called the Simulation-Based Stochastic Averaging Method to solve the discrepancy. In addition, Maruyama et al.~\cite{Maruyama2022_moment} proposed a method for estimating parametric rolling using the moment equation~\cite{Bover1978} (hereinafter referred to as the moment method). This method also allows to obtain some quantitative agreement with the result of MCS for the PDF of the parametric roll amplitude. On the other hand, there are relatively few cases in which these analytical methods for estimating roll motion have been validated in tank tests.
    %
    \par
    Moreover, estimation of the roll restoring variation is important in the theoretical estimation of parametric rolling. 
    In the studies of the theoretical estimation of parametric rollong in irregular waves by Maruyama et al.~\cite{Maruyama2022}\cite{Maruyama2022_moment}\cite{Maruyama2023_momentamp} mentioned above, the roll restoring variation was estimated by considering only the wave component based on the Froude-Krylov assumption under quasi-statically balancing heave and pitch in waves and introducing Grim's effective wave concept~\cite{Grim1961}.
    However, Hashimoto et al.~\cite{Hashimoto2004} suggest that the estimation method of the roll restoring variation considering only the wave component based on the Froude-Krylov assumption under quasi-statically balancing heave and pitch in waves might overestimate the risk of parametric rolling due to the large roll restoring variation during wave passage through the captive model test. Yu et al.~\cite{Yu2019} compare the degree of this overestimation using five models that estimate nonlinear restoring forces and Froude-Krylov forces. Furthermore, Hashimoto et al.~\cite{Hashimoto2006} measured the roll restoring variation in a captive model test in irregular waves and compared it with the results of the calculation of the roll restoring variation using Grim's effective wave. The results suggested that the accuracy of the roll restoring variation using Grim's effective wave remained a problem.
    %
\subsection{Object and Scope}
    The primary objective of this study is to validate theoretical estimation methods for the parametric roll amplitude in irregular waves and improve their accuracy. The contributions of this study are as follows:
    \begin{enumerate}
        \item Validation of theoretical estimation methods for parametric roll amplitude by comparison with corresponding experimental result,
        \item Suggestion for the method to improve the accuracy of estimation of the roll restoring variation in irregular waves, and
        \item Suggestion for the method to estimate the distribution of the maximum amplitude of parametric rolling in irregular waves.
    \end{enumerate}
    %
    In this study, we first conducted the model ship motion experiment in order to obtain the PDF of parametric roll amplitude in long-crested irregular waves in the towing tank of Osaka University.
    After that, the accuracy of the theoretical methods is validated by comparing the experimental results with the PDFs of the parametric roll amplitude calculated using three theoretical calculation methods; Roberts' stochastic averaging method~\cite{Roberts1982}, the energy-based stochastic averaging method~\cite{Dostal2011} (hereinafter referred to as ESAM), and the moment method~\cite{Maruyama2022_moment,Maruyama2023_momentamp}.
    Furthermore, we attempted to improve the accuracy of the theoretical estimation method by using the equation of motion reflecting the correction of the roll restoring variation based on the captive model test by Kono et al.
    \par 
    Extreme value theory is also widely used in risk assessment in various engineering fields. It would be significant to introduce the concept of extreme value theory to the risk assessment of parametric rolling motion. On the other hand, accurate estimation of the distribution of maximum values requires accurate estimation of the PDF of the parametric rolling amplitude down to the tail.
    Therefore, we pay particular attention to the behavior of the tail section of the PDF of the parametric roll amplitude in order to estimate extreme values. Finally, we propose a method for calculating the distribution of the maximum roll amplitude based on the theoretical method.
    %
%
%

\section{Subject ship}\label{sec:subject_ship}
    The subject ship is a C11 class post-Panamax container ship. In this study, a geometric similarity model with a scale of 1/100 is used for model tests at the towing tank of Osaka University. The main features of the model are shown in~ \Cref{tab:Principal}.
    %
    \begin{table}[h]
        \centering
        \caption{Principal Particulars of C11 at model and full scale.}
        \begin{tabular}{ccc}
        \hline Items & \multicolumn{2}{c} { Value } \\
        \hline \hline Scale & Model & Full \\
        $L_{\mathrm{pp}} \, \left( {\mathrm{m}} \right)$ & 2.62 & 262 \\
        $B \, \left( {\mathrm{m}} \right)$ & 0.4 & 40 \\
        $D \, \left( {\mathrm{m}} \right)$  & 0.2445 & 24.45 \\
        $d \, \left( {\mathrm{m}} \right)$ & 0.115 & 11.5 \\
        $W \, \left( {\mathrm{kg}} \right)$ & 67.247 & $6.7247 \times10^{7} $\\
        $C_{\mathrm{b}} $ & 0.56 & 0.56 \\
        $\mathrm{GM} \, \left( {\mathrm{m}} \right)$ & 0.019299 & 1.9299 \\
        $T_{\phi} \, \left( {\mathrm{s}} \right)$ & 2.44 & 24.4 \\
        \hline \hline
        \end{tabular}
        \label{tab:Principal}
    \end{table}
    The extinction coefficients used in the calculations were calculated from the results of free-rolling tests using the model in a water tank.

\section{Towing Tank Experiment}\label{towing_test}

    By conducting model ship motion experiments in long-crested irregular waves and obtaining the PDF of the roll amplitude, we validate the theoretical calculation method as described in~\Cref{sec:intro}. This section describes the experimental method used in this study.

\begin{figure}[tb]
        \centering
       \includegraphics[width=\linewidth]{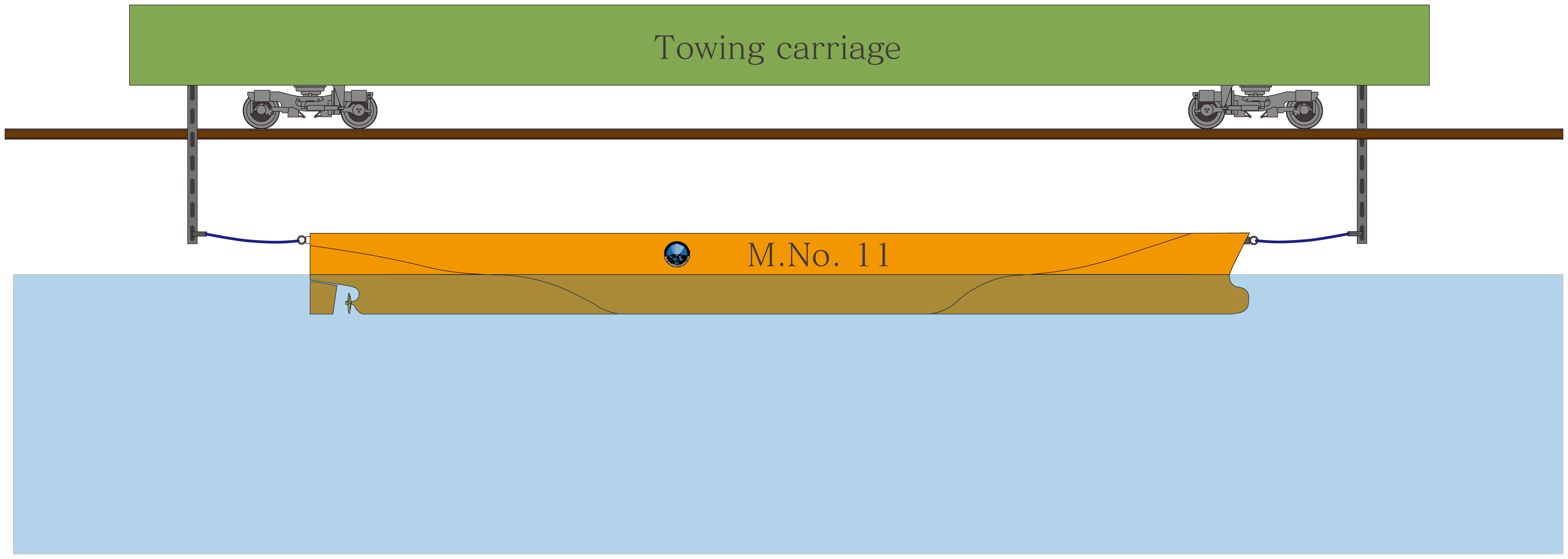}
        \caption{Schematic view of the experiment}
        \label{fig:situation_ship}
    \end{figure}

    The bow and stern of the model ship is connected to the towing dolly with a rubber strap to gently restrain the model ship. The tension of the rubber strap should be adjusted appropriately so as not to constrain the surge or roll motion too much. The stern rubber strap was not used in most of the experiments in head-wave conditions.
    The state of the model ship is shown in~\Cref{fig:situation_ship}. 
    The long-crested irregular head waves are generated using ITTC spectrum.
    The angle of occasional rolling of the model ship is measured by a fiber-optic gyro sensor mounted at the center of gravity of the model ship. Wave heights of irregular waves are measured by a wave height meter installed at the front of the model ship.
    \crefmiddleconjunction
    The experimental conditions for irregular waves are shown in~\Cref{tab:wave}.
    %
    \begin{table}[tb]
     \centering
     \caption{Wave condition at full scale}
    \begin{tabular}{ccc}
    \hline Items & \multicolumn{2}{c} { Value } \\
    \hline \hline 
    $H_{1/3} ~ \left( {\mathrm{m}} \right)$ & 5.0 & 7.0 \\
    $T_{01} ~ \left( {\mathrm{s}} \right)$ & 10.0 & 10.0 \\
    Num. of realization  & 24 & 24 \\
    \hline \hline
    \end{tabular}
    \label{tab:wave}
    \end{table}
    The producing time for each irregular wave is $300~\mathrm{s}$. Here, the data between $270~\mathrm{s}$ from $70~\mathrm{s}$ to $340~\mathrm{s}$ after the start of producing waves were considered for the analysis.
    %
    \begin{figure}[tb]
        \centering
        \includegraphics[width=\linewidth]{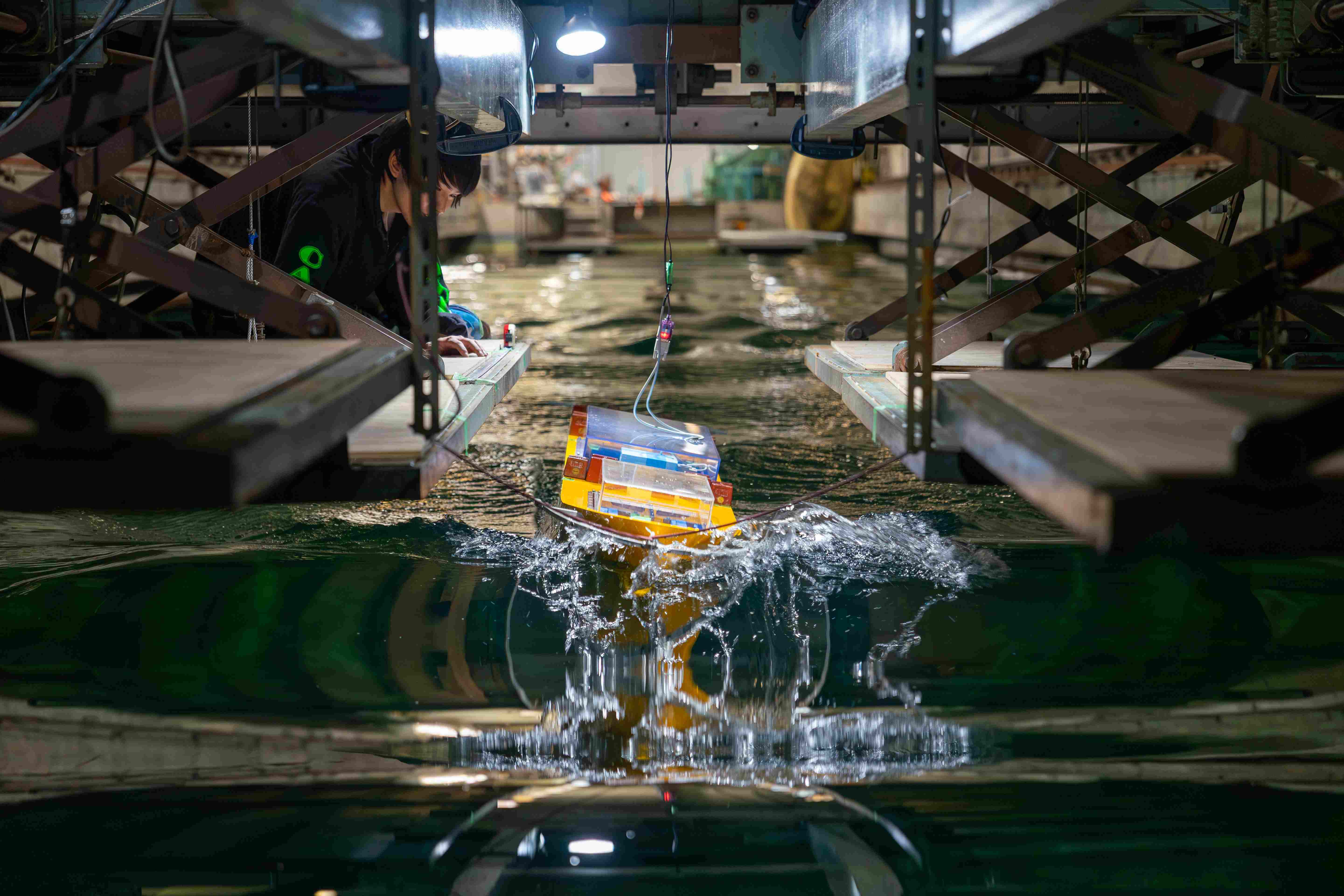}
        \caption{The situation of the experiment}
        \label{fig:situation_exp}
    \end{figure}

    \section{Experimental Results and Discussion}\label{sec:result_exp}
    \subsection{Analysis methods and theoretical estimation methods for roll amplitudes}
\begin{figure}[tb]
        \centering
        \includegraphics[width=\linewidth]{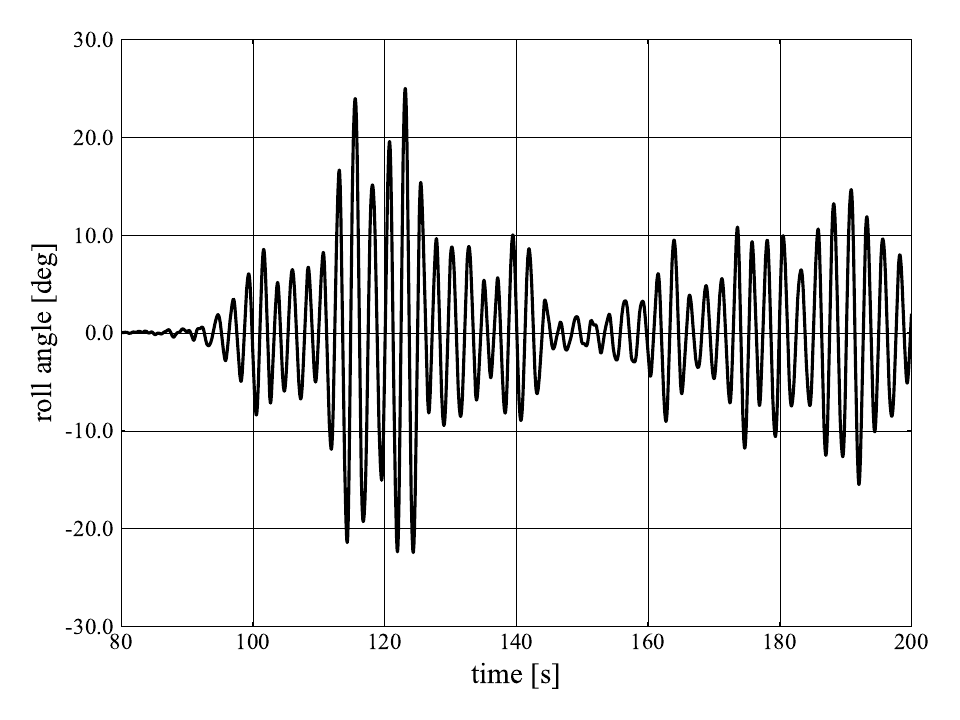}
        \caption{Time history of roll angle, $H_{1/3} = 7.0 ~\mathrm{m}$, $T_{01} = 10.0 ~ \mathrm{s}$}
        \label{fig:roll_sample}
    \end{figure}

    An example of a time series of the roll angle obtained in the experiment is shown in~\Cref{fig:roll_sample}.
    In this study, the following two methods of analyzing the roll amplitude are used.
    %
    \subsubsection{Experimental Zero Crossing method}
    In the time series data of the roll angle obtained in the experiment, 
    The time series data of roll angle is obtained by the experiment.
    The maximum value between every zero up crossing and zero down crossing and the minimum value between every zero down crossing and zero up crossing are recorded. The absolutes of them are the roll amplitudes.
    The PDF is calculated from the distribution of the obtained roll amplitudes.
    \label{sec:method_analysis_exp_zero_cross}
    %
    \subsubsection{Experimental Envelope method}
    The Hilbert transform of the time series of roll angles obtained in the experiment is used to create an envelope. Then, the absolute value of the envelope of the roll angle at each sampling point is recorded as the amplitude. The PDF is calculated from the distribution of the values of roll angles.
    \label{sec:method_analysisi_exp_env}
    %
    \subsection{Theoretical estimation methods}
    \label{sec:theo_method_para_roll_amp}
    Theoretical estimation of the amplitude of parametric rolling in irregular waves requires the estimation of the corresponding the roll restoring variation. Therefore, in this study, we introduce Grim's effective wave theory~\cite{Grim1961} for irregular waves. Moreover, considering only the wave component based on the Froude-Krylov assumption under quasi-statically balancing heave and pitch in waves, we obtain an equation relating the displacement of the wave at the center of the hull to the amount of GM variation~\cite{Maruyama2022}. In this way, we estimate GM variation in irregular waves. The specific procedure of this transformation method is described in detail in~\Cref{sec:appendix}~A. 
    Based on the roll restoring variation estimated in this way, we estimate the amplitude of parametric rolling in irregular waves using three theoretical methods, Roberts' stochastic averaging method \cite{Roberts1982}, ESAM~\cite{Dostal2011}, and the moment method~\cite{Maruyama2023_momentamp}.
    \vskip.5\baselineskip
    \subsection{Experimental results}
\begin{figure}[tb]
        \centering
       \includegraphics[width=\linewidth]{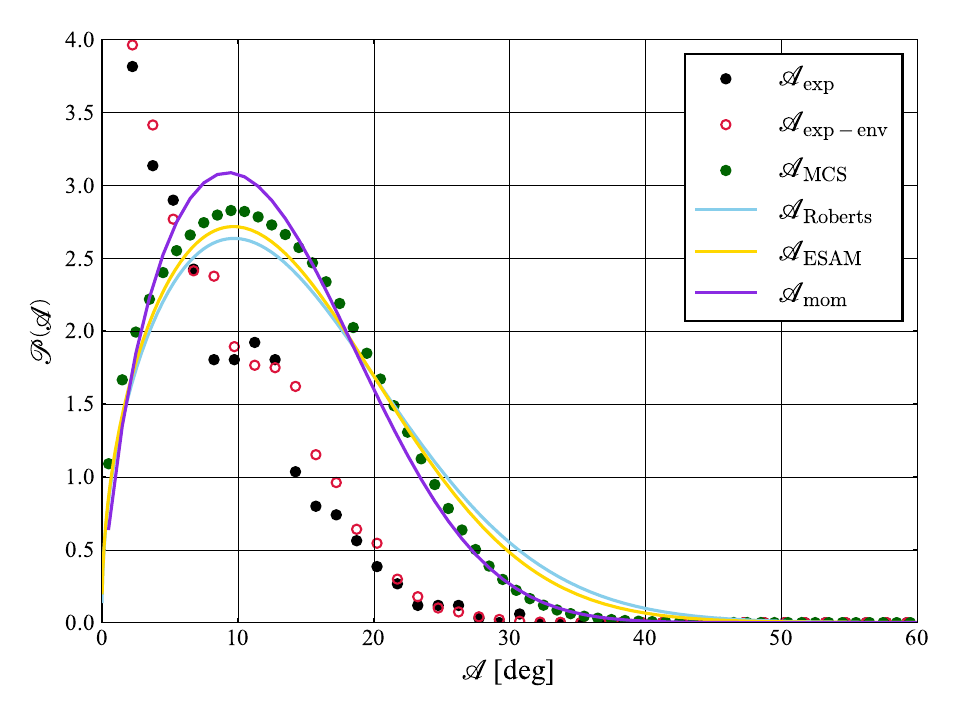}
        \caption{PDF of roll amplitude, $H_{1/3} = 5.0 ~ \mathrm{m}$, $T_{01} = 10.0 ~ \mathrm{s}$, linear scale}
        \label{fig:PDF_rollamp_H5_T10}
    \end{figure}

\begin{figure}[tb]
        \centering
       \includegraphics[width=\linewidth]{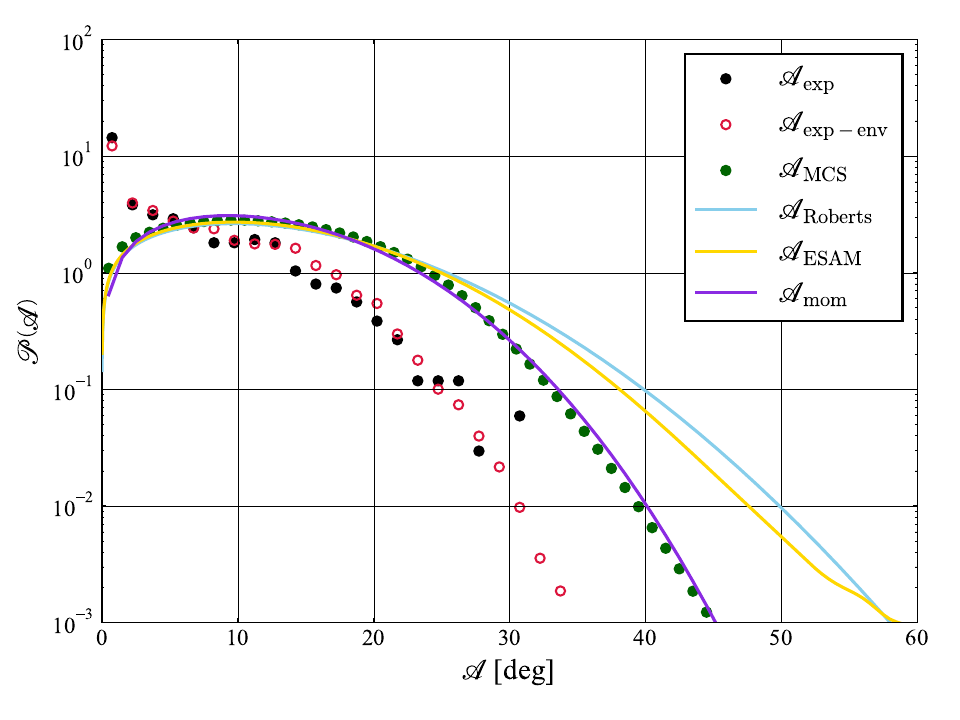}
        \caption{PDF of roll amplitude, $H_{1/3} = 5.0 ~ \mathrm{m}$, $T_{01} = 10.0 ~ \mathrm{s}$, logarithmic scale}
        \label{fig:PDF_rollamp_H5_T10_log}
    \end{figure}

\begin{figure}[tb]
        \centering
       \includegraphics[width=\linewidth]{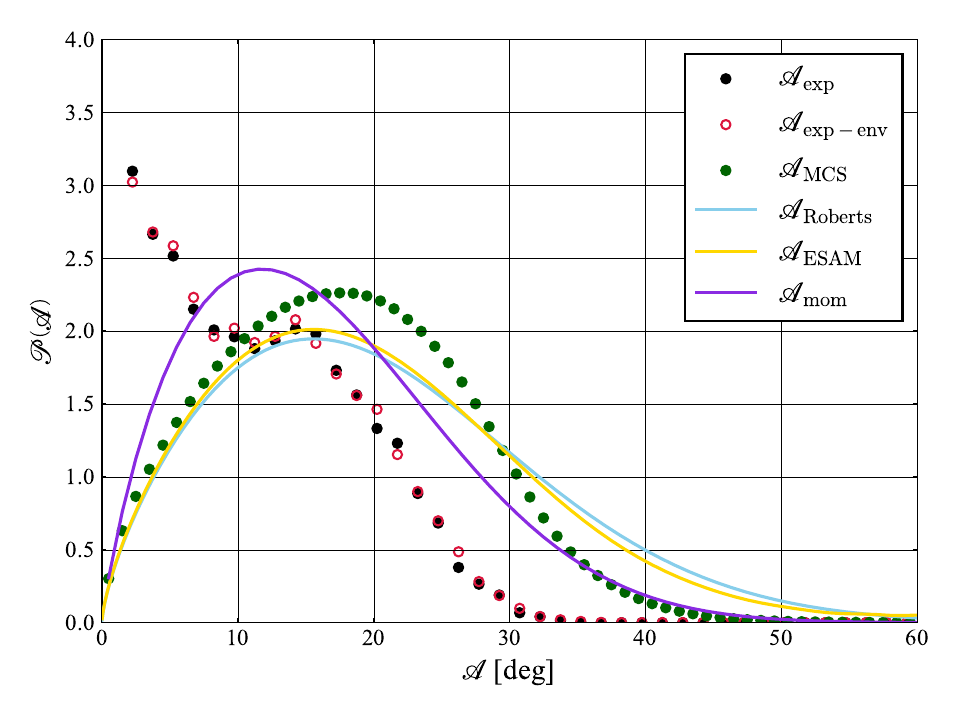}
        \caption{PDF of roll amplitude, $H_{1/3} = 7.0 ~ \mathrm{m}$, $T_{01} = 10.0 ~ \mathrm{s}$, linear scale}
        \label{fig:PDF_rollamp_H7_T10}
    \end{figure}

\begin{figure}[tb]
        \centering
       \includegraphics[width=\linewidth]{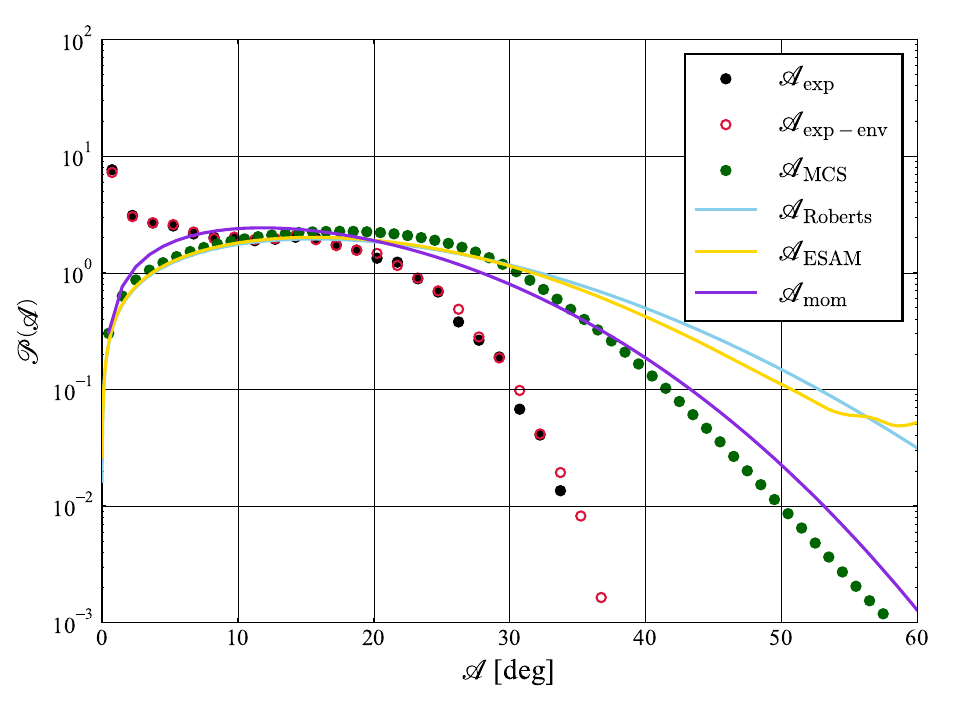}
        \caption{PDF of roll amplitude, $H_{1/3} = 7.0 ~ \mathrm{m}$, $T_{01} = 10.0 ~ \mathrm{s}$, logarithmic scale}
        \label{fig:PDF_rollamp_H7_T10_log}
    \end{figure}

    The results of comparing the PDFs of the roll amplitudes obtained from experiments and various theoretical calculation methods are shown in \Cref{fig:PDF_rollamp_H5_T10,fig:PDF_rollamp_H5_T10_log}, and \Cref{fig:PDF_rollamp_H7_T10,fig:PDF_rollamp_H7_T10_log}.
    Here, the black dots in the figure are the PDF calculated by the experimental zero crossing method described in~ \Cref{sec:method_analysis_exp_zero_cross} and the red dots are the PDF calculated by the experimental envelope method described in~ \Cref{sec:method_analysisi_exp_env}. The blue line is the PDF obtained by Roberts' stochastic averaging method ~\cite{Roberts1982}, the yellow line is the PDF obtained by ESAM~\cite{Dostal2011}, and the purple line is the PDF obtained by the moment method~\cite{Maruyama2023_momentamp}. The green dots are the PDF obtained by MCS. For MCS, the initial value of the roll angle was set to $5~\mathrm{deg}$, and the initial value of the roll velocity was set to $0~\mathrm{deg/s}$. The time for one trial of simulation was 3600 s. The number of trials was $10^4$ times.
    %
    %
    %
%
\subsection{Consideration} \label{sec:exp_discuss}
    The PDFs of the roll amplitudes obtained from the experiments differ quantitatively and qualitatively from the PDFs obtained from the simulation and from each theoretical prediction method.
    Around $0~\mathrm{deg}$, the simulation and theoretical results show that the PDF converges to $0$, while the experimental results are non-zero. The experimental values have a local maximum from $10~\mathrm{deg}$ to $20~\mathrm{deg}$, and then they are smaller than the theoretical values.
    In the experiments, an intermittent rolling behaviour was observed, whereas pure parametric rolling behavior resulted from the theoretical methods and the MCS of the equations of motion.
    In \Cref{fig:PDF_rollamp_H7_T10,fig:PDF_rollamp_H7_T10_log}, the form of the PDF from the experiments obtained here, which takes non-zero values around $0~\mathrm{deg}$ and has one more hump, is not consistent with the two forms of the theoretical PDFs presented in previous studies~\cite{Maruyama2023_momentamp,Maki2019}. We currently believe that the reason for such PDFs is the change of yaw angle during the experiment. Further investigation is necessary in the future.
    
    Next, we compare the three PDFs obtained by the theoretical method with that obtained by MCS. In \Cref{fig:PDF_rollamp_H7_T10}, the PDFs obtained by Roberts' stochastic averaging method and ESAM agree well with the result obtained by MCS in the amplitude range from $0~\mathrm{deg}$ to $10~\mathrm{deg}$.
    However, in the log scale in \Cref{fig:PDF_rollamp_H7_T10_log}, the PDF obtained by the moment method agrees better with the result of MCS in the tail (large amplitude part) of the PDF.
    As mentioned in~\Cref{sec:intro}, the estimation of the maximum roll amplitude requires an accurate estimation of the PDF of the roll amplitude down to the tail part. Therefore, when consider the PDF of the maximum roll amplitude in~\Cref{sec:PDF_maximum} using theoretical methods, the results obtained by the moment method, which are successful in quantitatively estimating the tail, are considered more suitable for use in~\Cref{eq:func_PDF_maximum}.
    %

\section{Correction of GM variation}\label{sec:gm_variation}
    As shown in~\Cref{fig:PDF_rollamp_H7_T10,fig:PDF_rollamp_H7_T10_log}, there was a discrepancy between the experimental PDF and the PDF obtained by MCS and each theoretical calculation. We attempt to reduce this discrepancy in order to improve the estimation accuracy of the PDF of the parametric roll amplitude.
    In this study, GM variation is calculated by the equation relating wave displacement and GM variation.  We consider only the wave component based on the Froude-Krylov assumption under quasi-statically balancing heave and pitch in waves, and introduce Grim's effective wave theory~\cite{Grim1961} for irregular waves as explained in~\Cref{sec:theo_method_para_roll_amp} \cite{Maruyama2022}. The specific calculation procedure is described in~\Cref{sec:appendix} A.
    However, the results of captive model tests in regular following waves by Hashimoto et al.~\cite{Hashimoto2004} suggests that the estimation method of the roll restoring variation considering only the wave component based on the Froude-Krylov assumption under quasi-statically balancing heave and pitch in waves might overestimate the risk of parametric rolling due to the large roll restoring variation during wave passage. Furthermore, Hashimoto et al.~\cite{Hashimoto2006} measured the roll restoring variation in a captive model test in irregular waves and compared it with the results of the calculation of the roll restoring variation using Grim's effective wave. The results suggest that the accuracy of the roll restoring variation using Grim's effective wave remain a problem. Therefore, we consider the accuracy of the estimation of the roll restoring variation to be one of the reasons for the discrepancy between the experimental and theoretical results for the PDF of the parametric roll amplitude in irregular waves. Hence, captive model tests were conducted to evaluate the estimation accuracy of the theoretical equation for GM variation.
    \subsection{Captive model test}\label{subsec:captive_test} 
        The experimental method for the captive model testing is explained in the following.
        The model ship was free to heave and pitch, and was attached to the towing vehicle via a quarter force gauge. The fore-and-aft force ($F_X$), lateral force ($F_Y$), turning moment ($N$), lateral tilting moment ($K$), heaving, and pitching on the hull were measured. In addition, irregular waves were generated by a flap wave generator at the end of the tank, and the generated irregular wave forms were measured using a capacitance type wave height meter installed between the wave generator and the model ship.
        Three wave conditions were used in this experiment, see~\Cref{tab:wave_captive}. For each wave condition, the number of realizations is $10$. 
        \begin{table}[tb]
            \centering
            \caption{Wave condition at full scale on the captive model test}
            \begin{tabular}{cccc}
            \hline Items & \multicolumn{3}{c} { Value } \\
            \hline \hline 
            $H_{1/3} ~ \left( {\mathrm{m}} \right)$ & 5.0 & 5.0 & 7.0 \\
            $T_{01} ~ \left( {\mathrm{s}} \right)$ & 10.0 & 12.37 & 10.0 \\
            Num. of realizations  & 10 & 10 & 10 \\
            \hline \hline
            \end{tabular}
            \label{tab:wave_captive}
        \end{table}
        In the experiment of this study, the fixed roll angles $\phi$ were 0, 5, 10, and 15 degrees. The GM variation $\Delta\mathrm{GM}$ was obtained from the difference between the roll moment at $\phi = 0~\mathrm{deg}$.
        \begin{figure}[tb]
            \centering
            \includegraphics[width=\linewidth]{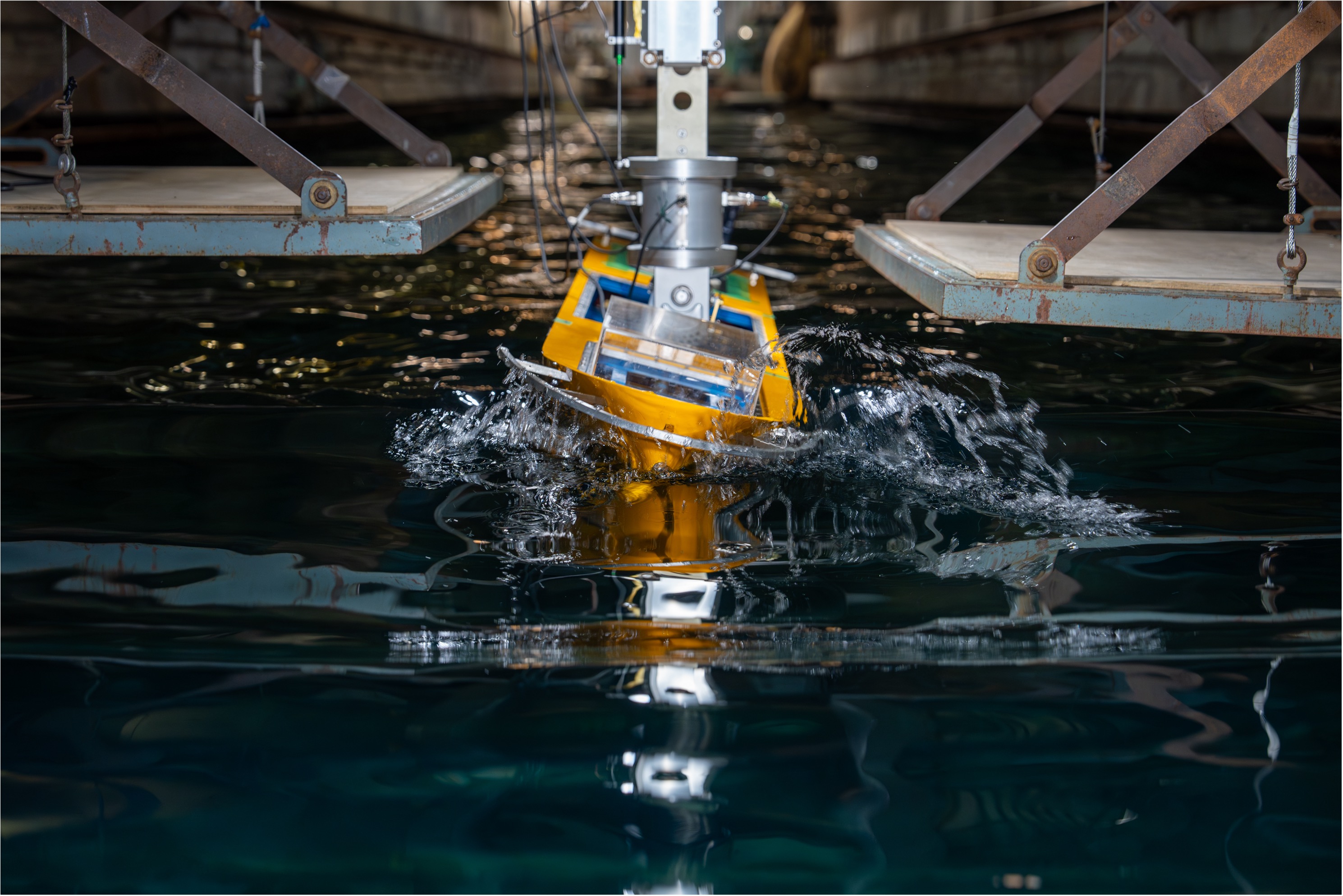}
            \caption{Condition during captive model test}
            \label{fig:situation_captive}
        \end{figure}

    \subsection{Experimental result and Method of correcting GM variation}
    \label{sec:method_GM_correction}
    
\begin{figure}[tb]
            \centering
            \includegraphics[width=0.8\linewidth]{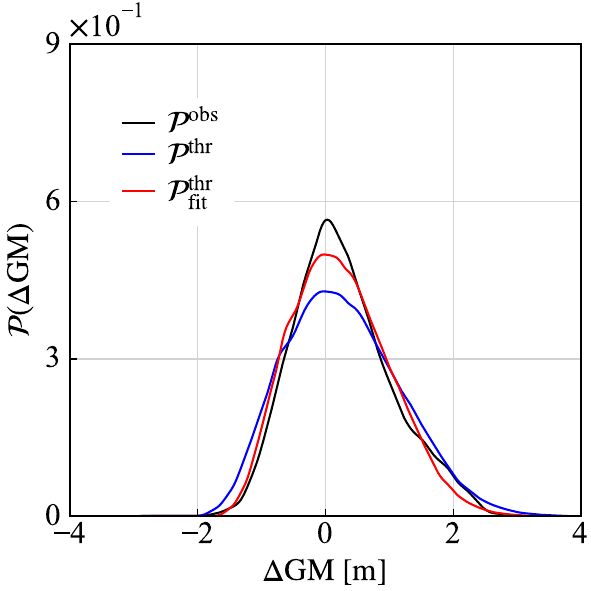}
            \caption{$\mathrm{PDF}$ of $\Delta\mathrm{GM}~$with$~\mathrm{H}_{1/3}=0.07~\mathrm{m}$,$~\mathrm{T}_{01}=1.0~\mathrm{s}$, linear scale}
            \label{fig:pdf for exp and thr1}
        \end{figure}

\begin{figure}[tb]
            \centering
            \includegraphics[width=0.8\linewidth]{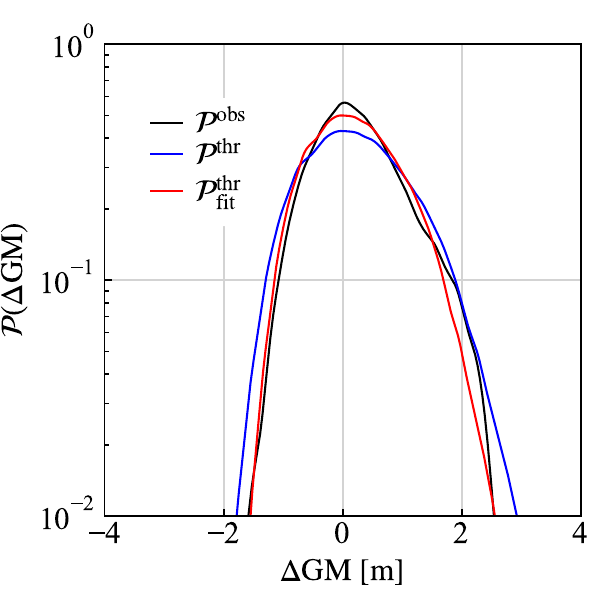}
            \caption{$\mathrm{PDF}$ of $\Delta\mathrm{GM}~$with$~\mathrm{H}_{1/3}=0.07~\mathrm{m}$,$~\mathrm{T}_{01}=1.0~\mathrm{s}$, logarithmic scale)}
            \label{fig:pdf for exp and thr2}
        \end{figure}

        The theoretical $\mathrm{PDF}$ is calculated based on $\Delta\mathrm{GM}(t)$ created in~\Cref{sec:appendix} A.
        The experimental $\mathrm{PDF}$ and theoretical PDFs are shown in~\Cref{fig:pdf for exp and thr1,fig:pdf for exp and thr2}. The black line is the experimental value and the blue line is the theoretical value based on the relation~\Cref{eq:non-memory}. By comparison, the theoretical PDF has a wider range of $\Delta \mathrm{GM}$ than the experimental PDF. Therefore, it is found that the conventional theoretical PDFs are overestimated compared to the experimental PDFs. Now, assuming that $\Delta\mathrm{GM}$ is normally distributed, we correct~\Cref{eq:non-memory} as~\Cref{eq:non-memory new} using the ratio of the standard deviation $\sigma^{\mathrm{obs}}_\mathrm{GM}$ of the experimental values to the standard deviation $\sigma^{\mathrm{thr}}_\mathrm{GM}$ of the theoretical values. In~\Cref{fig:pdf for exp and thr1,fig:pdf for exp and thr2}, the red line is the theoretical value based on~\Cref{eq:non-memory new}.
        %
        \begin{equation}
            \Delta \mathrm{GM}(\zeta_{\mathrm{mid}}) = \frac{\sigma^{\mathrm{obs}}_\mathrm{GM}}{\sigma^{\mathrm{thr}}_\mathrm{GM}}\sum_{k=0}^{6} \mathrm{C}_\mathrm{k}\zeta_{\mathrm{mid}}^{\mathrm{k}}\label{eq:non-memory new}
        \end{equation}
        %
    %
    \subsection{Calculation results of parametric rolling amplitude}
\begin{figure}[tb]
            \centering
           \includegraphics[width=\linewidth]{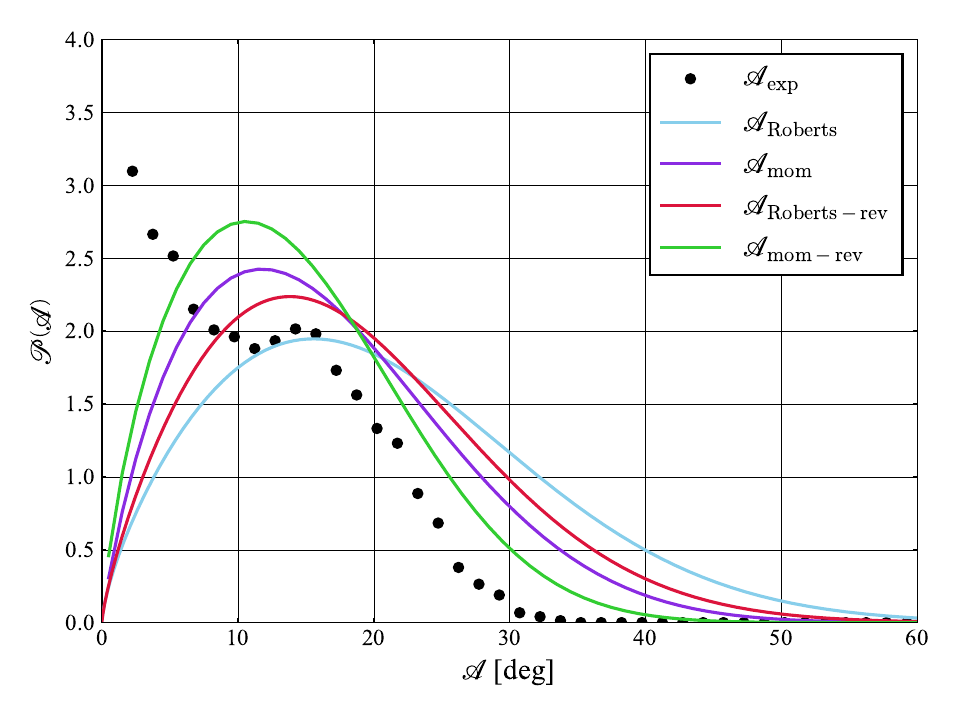}
            \caption{Comparison of the PDFs of the roll amplitude before and after correction, $H_{1/3} = 7.0 ~ \mathrm{m}$, $T_{01} = 10.0 ~ \mathrm{s}$, linear scale}
            \label{fig:GM_adjusted}
        \end{figure}

\begin{figure}[tb]
            \centering
            \includegraphics[width=\linewidth]{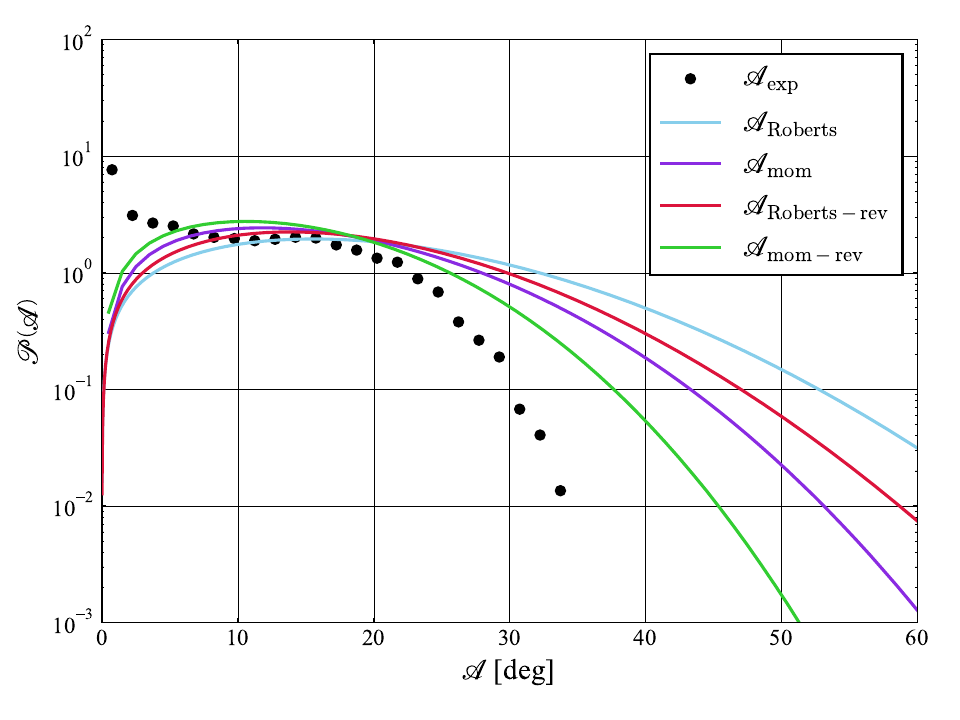}
            \caption{Comparison of PDF of roll amplitude before and after correction, $H_{1/3} = 7.0 ~ \mathrm{m}$, $T_{01} = 10.0 ~ \mathrm{s}$, logarithmic scale}
            \label{fig:GM_adjusted_log}
        \end{figure}

        The conventional equation of motion for one-degree-of-freedom roll motion is shown in~\Cref{eq:equation_motion}, where $\phi$ is the roll angle. $b_1$, $b_2$, and $b_3$ are the linear, quadratic, and cubic damping coefficients divided by $I_{xx}$, where $I_{xx}$ is the moment of inertia in roll including the corresponding added moment of inertia. The coefficients of the polynomial approximation of the GZ curve are $a_{i}~(i=1,~3,~5,~7,~\mathrm{and}~9)$, ${M}_{\mathrm{w}}({t})$ is the moment related to waves. The GM variation term is~${P}({t})$, which can be expressed by \Cref{eq:def_P}. Moreover, $\omega_0$ is natural roll frequency, and $\mathrm{GM}_0$ is the metacentric height in still water.
        \begin{equation}
            \begin{split}
            &\frac{\mathrm{d}^2 \phi}{\mathrm{d} {t}^2}+{b}_1 \frac{\mathrm{d} \phi}{\mathrm{d} {t}}+{b}_2 \frac{\mathrm{d} \phi}{\mathrm{d} {t}}\left|\frac{\mathrm{d} \phi}{\mathrm{d} {t}}\right|+{b}_3\left(\frac{\mathrm{d} \phi}{\mathrm{d} {t}}\right)^3\\&+\sum_{n=1}^5 {a}_{2 n-1} \phi^{2 n-1}+{P}({t}) \phi={M}_{\mathrm{w}}({t})
            \end{split}
            \label{eq:equation_motion}
        \end{equation}
        \begin{equation}
            {P}({t})=\frac{\omega_0^2}{\textup{GM}_0} {\textup{GM}}({t})
            \label{eq:def_P}
        \end{equation}
        According to~\Cref{sec:method_GM_correction}, we multiplied the modification coefficient ${\sigma^{\mathrm{obs}}_\mathrm{GM}}/{\sigma^{\mathrm{thr}}_\mathrm{GM}}$ by ${P}({t})$. The final equation of motion is given by
        %
        \begin{equation}
            \begin{split}
                &\frac{\mathrm{d}^2 \phi}{\mathrm{d} {t}^2}+{b}_1 \frac{\mathrm{d} \phi}{\mathrm{d} {t}}+{b}_2 \frac{\mathrm{d} \phi}{\mathrm{d} \bar{t}}\left|\frac{\mathrm{d} \phi}{\mathrm{d} {t}}\right|+{b}_3\left(\frac{\mathrm{d} \phi}{\mathrm{d} {t}}\right)^3\\&+\sum_{n=1}^5 {a}_{2 n-1} \phi^{2 n-1}+\frac{\sigma^{\mathrm{obs}}_\mathrm{GM}}{\sigma^{\mathrm{thr}}_\mathrm{GM}}{P}({t}) \phi={M}_{\mathrm{w}}({t}).
            \end{split}
            \label{eq:equation_motion_rev}
        \end{equation}
        Applying Roberts stochastic averaging method~\cite{Roberts1982} and the moment method~\cite{Maruyama2023_momentamp} to the equations of motion ~\Cref{eq:equation_motion} and \Cref{eq:equation_motion_rev} before and after correction, respectively, the PDFs of the roll amplitude are obtained. These theoretical PDFs are compared with the PDF obtained from the experiment in~\Cref{fig:GM_adjusted,fig:GM_adjusted_log}. The black dots denote the experimental result, the blue line and the red line are the theoretical results obtained by Roberts' stochastic averaging method before and after correction, and the purple line and the green line are the theoretical results obtained by the moment method before and after correction.


        The results of \Cref{fig:GM_adjusted,fig:GM_adjusted_log} show that the calculation results after the correction of the GM variation are closer to the experimental results than before the correction, and the accuracy of the quantitative estimation of the roll amplitude is successfully improved. However, there is still a large difference between the experimental and theoretical results, mainly due to the discrepancy in the region of small roll angles. Our theoretical method using the corrected equation of motion could not completely represent an intermittent rolling behavior as in the experiments. Therefore, further improvement of used model equations and the theoretical calculation methods is desirable.

\section{PDF of the maximum roll amplitude}
    \label{sec:PDF_maximum}
    In order to estimate the distribution of the maximum values of the parametric roll amplitude, a theoretical expression for the PDF of the maximum roll amplitude is derived. The PDF of the maximum roll amplitude is then estimated based on the PDF obtained by the moment method, which is in quantitative agreement with the MCS results in~\Cref{sec:result_exp}. Moreover, the PDF estimated by the theoretical method is compared with the results of MCS in order to confirm the accuracy of the estimation of the PDF of the maximum roll amplitude.
    \subsection{Derivation of a theoretical formula to estimate the PDF of the maximum roll amplitude}
        Suppose now that $N_0$ amplitudes are extracted from the population of roll amplitudes and the highest value among them is $\mathcal{A}_{\mathrm{M}}$, where $\mathcal{A}_{\mathrm{M}}$ is a dimensionless value. Denoting the PDF of $\mathcal{A}_{\mathrm{M}}$ as $\mathcal{P}^*\left(\mathcal{A}_{\mathrm{M}}\right)$, the probability that the maximum value of the amplitude $\mathcal{A}$ is in $\lbrack \mathcal{A}_{\mathrm{M}},~\mathcal{A}_{\mathrm{M}}+\mathrm{d} \mathcal{A}_{\mathrm{M}}\rbrack$ is $\mathcal{P}^*\left(\mathcal{A}_{\mathrm{M}}\right) \mathrm{d} \mathcal{A}_{\mathrm{M}}$.
        \par
        This is the probability that only one of the $N_0$ amplitudes is between $\mathcal{A}_{\mathrm{M}}$ and $\mathcal{A}_{\mathrm{M}}+\mathrm{d} \mathcal{A}_{\mathrm{M}}$ and the remaining $\left(N_0 - 1\right)$ amplitudes are less than $\mathcal{A}_{\mathrm{M}}$. Thus $\mathcal{P}^*\left(\mathcal{A}_{\mathrm{M}}\right)$ can be expressed as follows \Cref{eq:func_PDF_maximum}~\cite{Goda2000}
        %
        \begin{equation}
            \begin{split}           \mathcal{P}^*\left(\mathcal{A}_{\mathrm{M}}\right) \mathrm{d} \mathcal{A}_{\mathrm{M}} 
                =& N_0 {\left[ 1 - \int_{\mathcal{A}_{\mathrm{M}}}^{\infty} \mathcal{P}\left(\mathcal{A}\right) \mathrm{d}\mathcal{A}\right]}^{N_0 - 1} \\&\mathcal{P}\left(\mathcal{A}_{\mathrm{M}}\right) \mathrm{d} \mathcal{A}_{\mathrm{M}}
            \end{split}  
            \label{eq:func_PDF_maximum}
        \end{equation}
        where $\mathcal{P}\left(\mathcal{A}_{\mathrm{M}}\right)$ is the value of PDF when the amplitude $\mathcal{A}$ is $\mathcal{A}_{\mathrm{M}}$.

        From the above, $\mathcal{P}^*\left(\mathcal{A}_{\mathrm{M}}\right)$, the PDF of the maximum roll amplitude, is obtained.
        %
        \par
        Also, when $N_0$ is large enough, i.e. $N_0 \rightarrow \infty$, then 
        %
        \begin{equation}
            \xi = N_0 \int_{\mathcal{A}_{\mathrm{M}}}^{\infty} \mathcal{P}\left(\mathcal{A}\right) \mathrm{d}\mathcal{A}
        \end{equation}
        %
        \begin{equation}
            \begin{split}
                \lim_{N_0 \to \infty} {\left[ 1 - \int_{\mathcal{A}_{\mathrm{M}}}^{\infty} \mathcal{P}\left(\mathcal{A}\right) \mathrm{d}\mathcal{A}\right]}^{N_0} 
                &= \lim_{N_0 \to \infty} {\left[ 1 - \frac{\xi}{N_0} \right]}^{N_0} \\
                &= e^{-\xi}
            \end{split}
        \end{equation}
        Therefore, \Cref{eq:func_PDF_maximum} can be expressed in terms of \Cref{eq:func_PDF_maximum_rev}.
        %
        \begin{equation}
            \begin{split}
                &\mathcal{P}^*\left(\mathcal{A}_{\mathrm{M}}\right) \mathrm{d} \mathcal{A}_{\mathrm{M}} \\
                =& \xi  e^{-\xi} {\left[ \int_{\mathcal{A}_{\mathrm{M}}}^{\infty} \mathcal{P}\left(\mathcal{A}\right) \mathrm{d}\mathcal{A} \left\{ 1 - \int_{\mathcal{A}_{\mathrm{M}}}^{\infty} \mathcal{P}\left(\mathcal{A}\right) \mathrm{d}\mathcal{A} \right\} \right]}^{- 1} \\ &\mathcal{P}\left(\mathcal{A}_{\mathrm{M}}\right) \mathrm{d} \mathcal{A}_{\mathrm{M}}
                \label{eq:func_PDF_maximum_rev}
            \end{split}
        \end{equation}

    \subsection{Calculation Method}
        The specific procedure for the calculation of the PDF of the maximum roll amplitude using MCS and  theoretical method based on \Cref{eq:func_PDF_maximum} is explained as follows.
        The initial value of the roll angle on MCS is set to $5~\mathrm{deg}$, and
        the first $N_0$ amplitudes are extracted from the time series data of the roll angle after $500~\mathrm{s}$ from the start of the simulation. The maximum value among the $N_0$ amplitudes is recorded as the maximum value in one trial. By repeating this $N$ times,
        $N$ maximum values are obtained in total. Based on these $N$ maxima, 
        the PDF of the maximum roll amplitude is calculated.
        Then, the shapes of each PDF with $N_0$ as the variable are compared.
        \par
        On the other hand, $\mathcal{P}\left(\mathcal{A}\right)$, the PDF of the roll amplitude, is calculated using the moment method and Roberts' stochastic averaging method. Using this, the PDF of the maximum roll amplitude is theoretically derived by using \Cref{eq:func_PDF_maximum}. Then, the PDF of the maximum roll amplitude calculated by the theoretical method is compared with that obtained from the MCS results.
        %

        \subsection{Numerical Results}
\begin{figure}[htb]
                \centering
                \includegraphics[width=\linewidth]{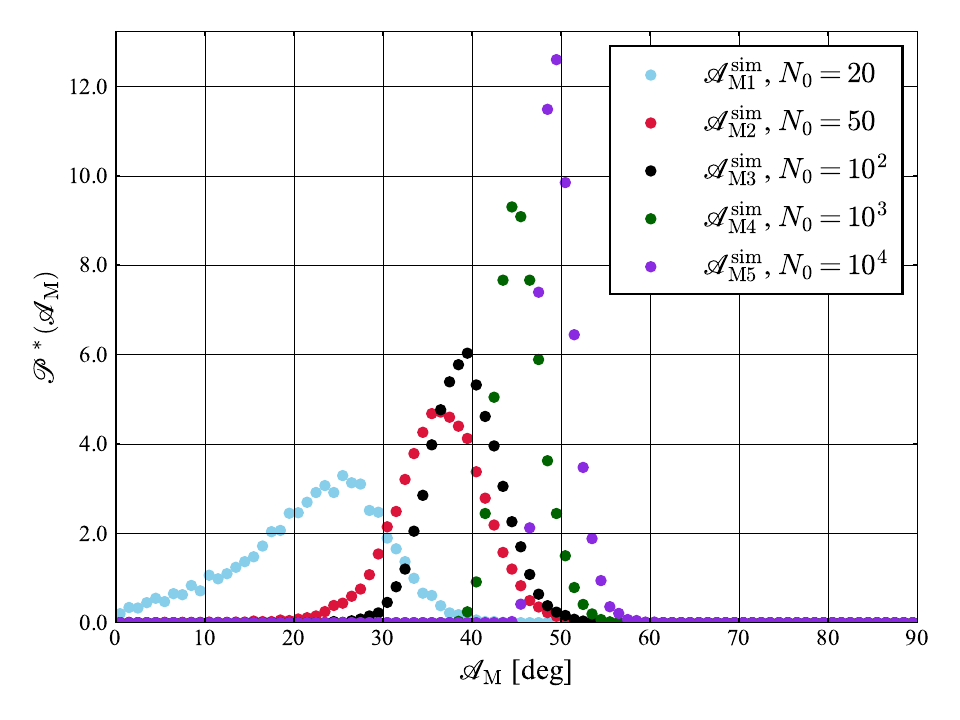}
                \caption{PDF of the maximum of roll amplitudes obtained by MCS, $H_{1/3} = 7.0 ~ \mathrm{m}$, $T_{01} = 10.0 ~ \mathrm{s}$, linear scale}
                \label{fig:PDF_max_everyN0}
            \end{figure}

\begin{figure}[htb]
                \centering
                \includegraphics[width=\linewidth]{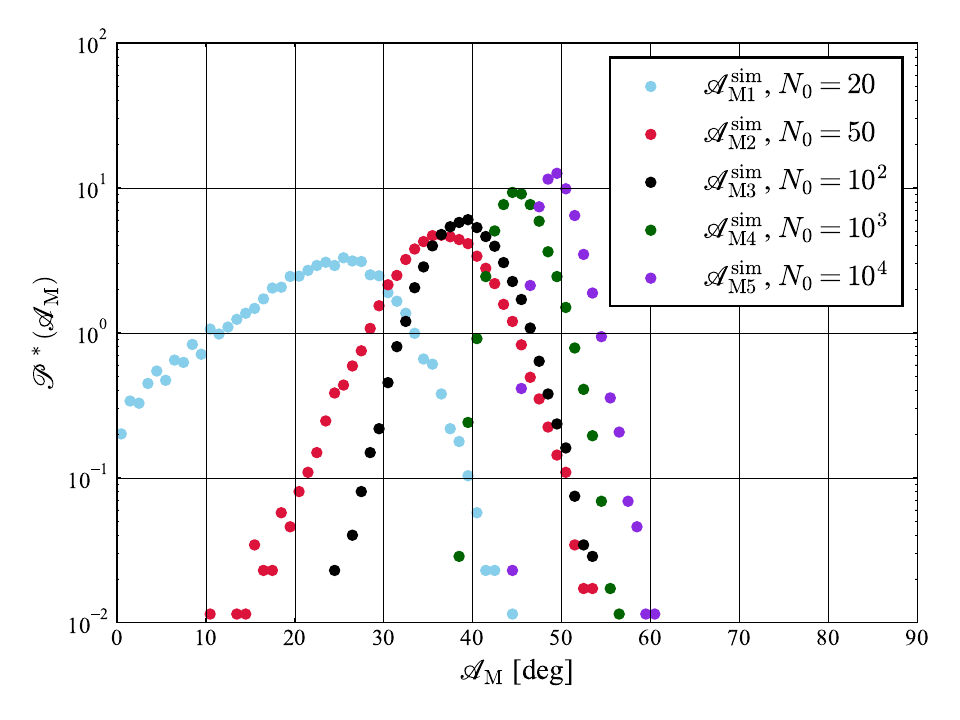}
                \caption{PDF of the maximum of roll amplitudes obtained by MCS, $H_{1/3} = 7.0 ~ \mathrm{m}$, $T_{01} = 10.0 ~ \mathrm{s}$, logarithmic scale}
                \label{fig:PDF_max_everyN0_log}
            \end{figure}

\begin{figure}[tb]
                \centering
                \includegraphics[width=\linewidth]{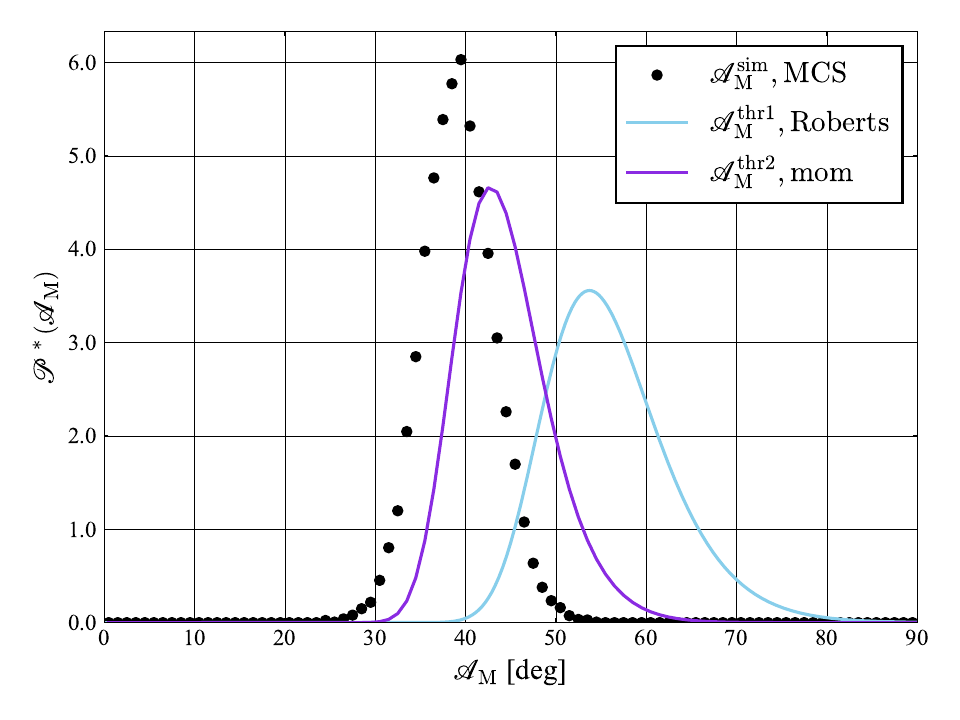}
                \caption{Comparison of PDF of the maximum of roll amplitudes obtained theoretical method and MCS, $H_{1/3} = 7.0 ~ \mathrm{m}$, $T_{01} = 10.0 ~\mathrm{s}$, $N_0 = 10^2$, linear scale}
                \label{fig:PDF_max_N0_100}
            \end{figure}

\begin{figure}[tb]
                \centering
                \includegraphics[width=\linewidth]{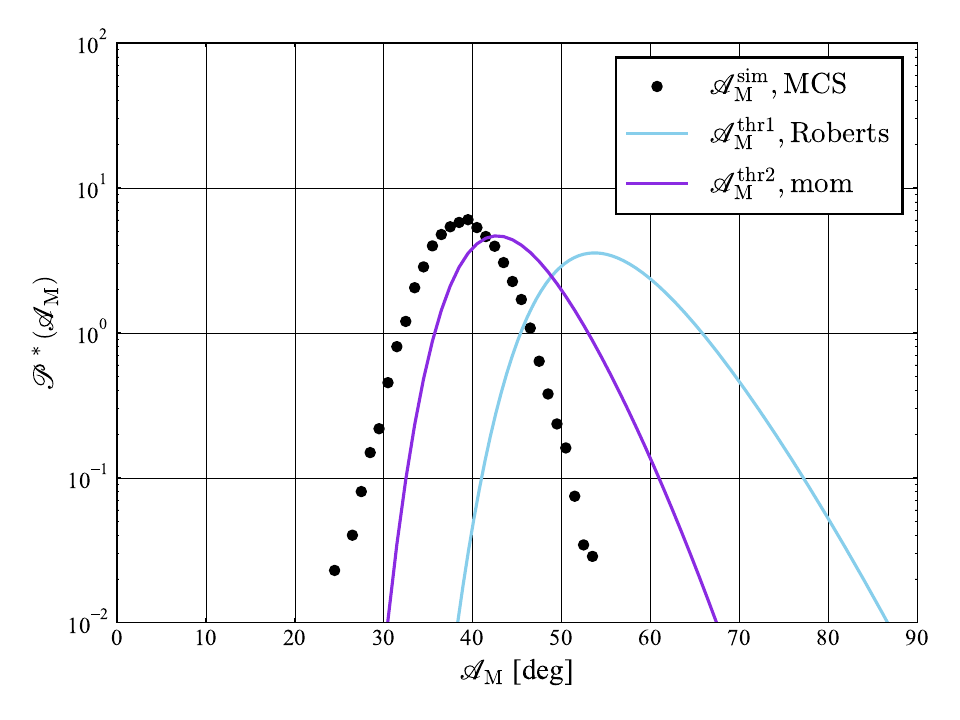}
                \caption{Comparison of PDF of the maximum roll amplitudes obtained by the theoretical methods and MCS, $H_{1/3} = 7.0 ~ \mathrm{m}$, $T_{01} = 10.0 ~\mathrm{s}$, $N_0 = 10^2$, logarithmic scale}
                \label{fig:PDF_max_N0_100_log}
            \end{figure}

\begin{figure}[tb]
                \centering
                \includegraphics[width=\linewidth]{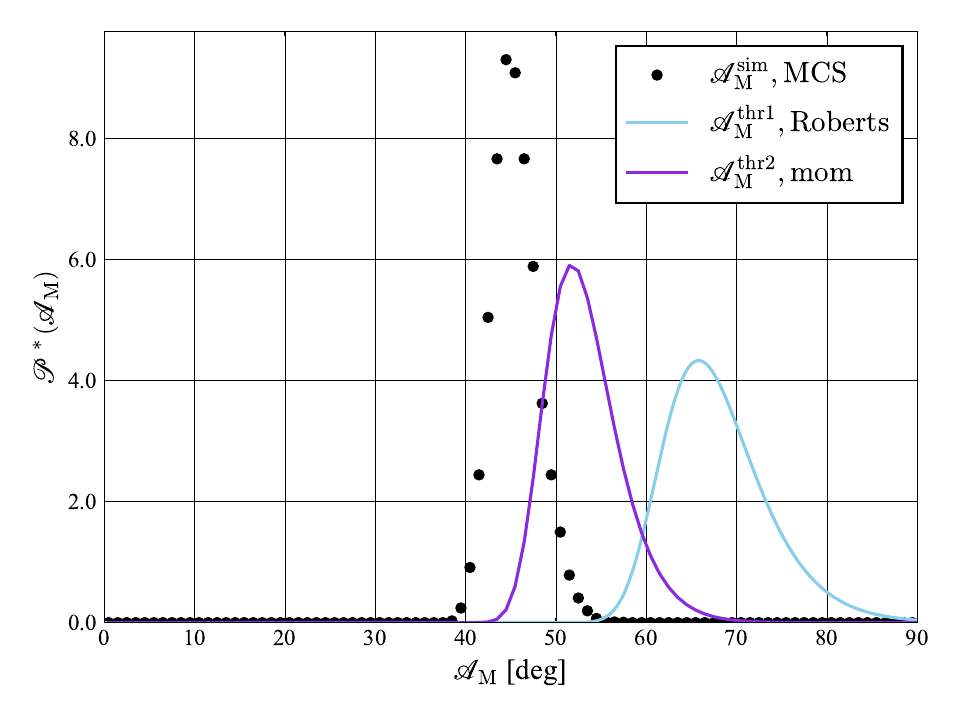}
                \caption{Comparison of PDF of the maximum roll amplitudes obtained by the theoretical methods and MCS, $H_{1/3} = 7.0 ~ \mathrm{m}$, $T_{01} = 10.0 ~\mathrm{s}$, $N_0 = 10^3$, linear scale}
                \label{fig:PDF_max_N0_1000}
            \end{figure}

\begin{figure}[tb]
                \centering
                \includegraphics[width=\linewidth]{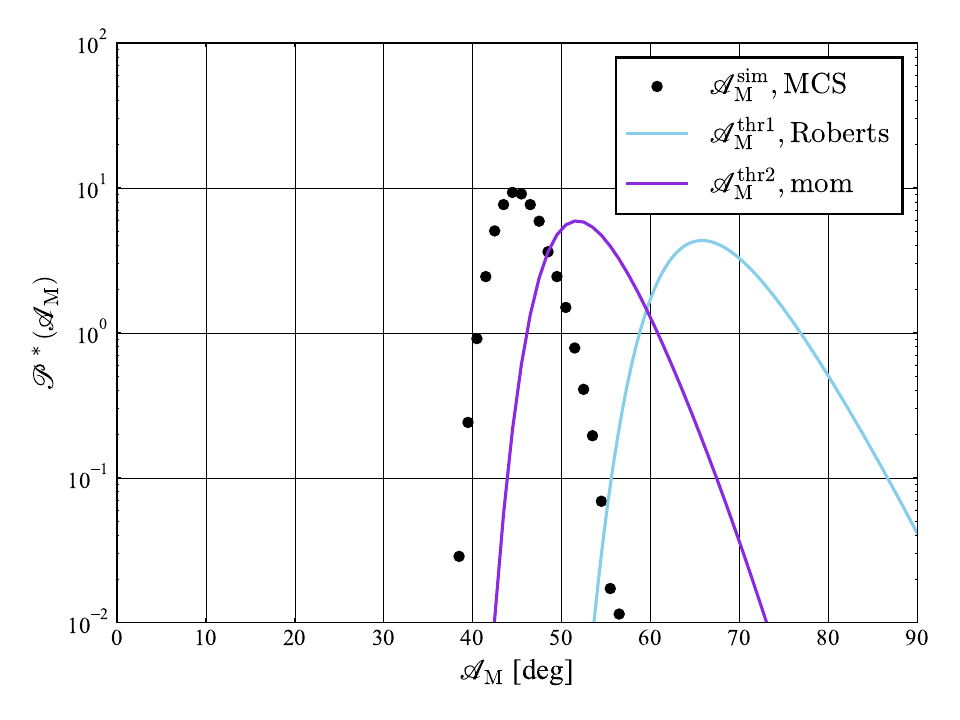}
                \caption{Comparison of PDF of the maximum roll amplitudes obtained by the theoretical methods and MCS, $H_{1/3} = 7.0 ~ \mathrm{m}$, $T_{01} = 10.0 ~\mathrm{s}$, $N_0 = 10^3$, logarithmic scale}
                \label{fig:PDF_max_N0_1000_log}
            \end{figure}

\begin{figure}[tb]
                \centering
                \includegraphics[width=\linewidth]{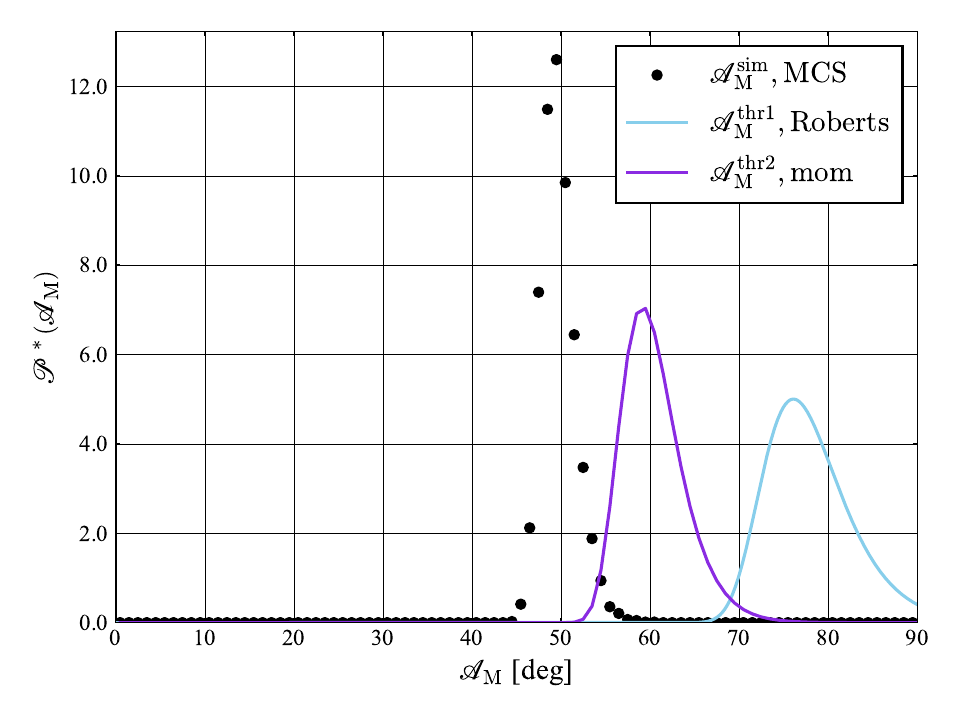}
                \caption{Comparison of PDF of the maximum roll amplitudes obtained by the theoretical methods and MCS, $H_{1/3} = 7.0 ~ \mathrm{m}$, $T_{01} = 10.0 ~\mathrm{s}$, $N_0 = 10^4$, linear scale}
                \label{fig:PDF_max_N0_10000}
            \end{figure}

\begin{figure}[tb]
                \centering
                \includegraphics[width=\linewidth]{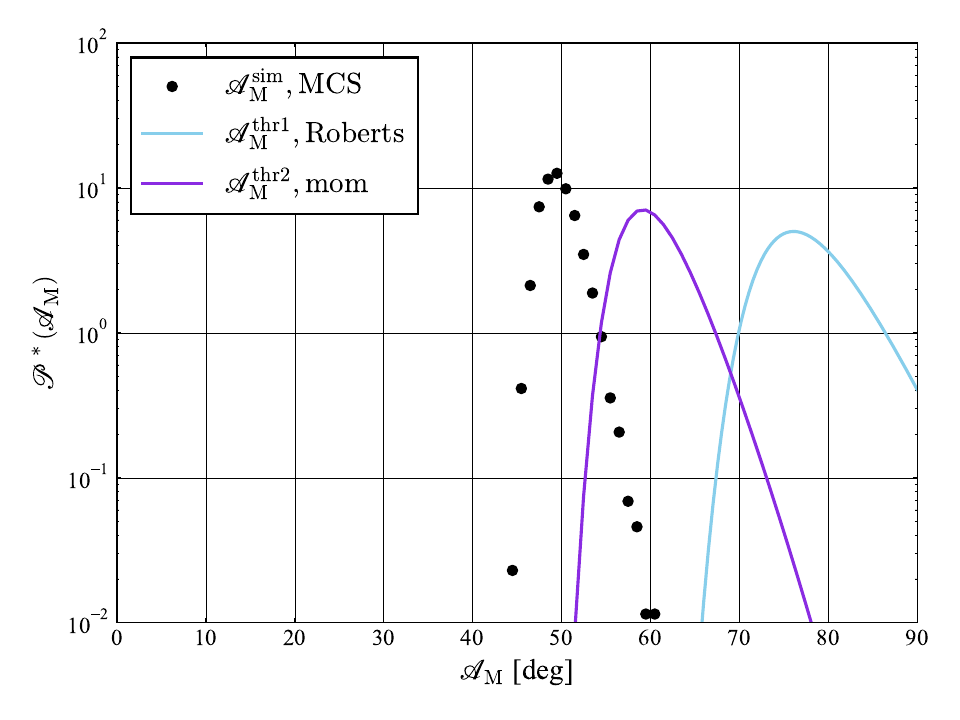}
                \caption{Comparison of PDF of the maximum roll amplitudes obtained by the theoretical methods and MCS, $H_{1/3} = 7.0 ~ \mathrm{m}$, $T_{01} = 10.0 ~\mathrm{s}$, $N_0 = 10^4$, logarithmic scale}
                \label{fig:PDF_max_N0_10000_log}
            \end{figure}

            With $N_0=20,~50,~10^2,~10^3,~10^4$, the PDFs of the maximum roll amplitude calculated by MCS are shown in \Cref{fig:PDF_max_everyN0,fig:PDF_max_everyN0_log}.
            When $N_0=20,~50,~10^2,~10^3$, the number of MCS trials $N$ is $10^4$.
            When $N_0=10^4$, $N$ is $5 \times 10^3$.
            %
            The results calculated by \Cref{eq:func_PDF_maximum} using the moment method and Roberts' stochastic averaging method are compared with that obtained from MCS and are shown in \Cref{fig:PDF_max_N0_100,fig:PDF_max_N0_100_log}, and \Cref{fig:PDF_max_N0_1000,fig:PDF_max_N0_1000_log}, and \Cref{fig:PDF_max_N0_10000,fig:PDF_max_N0_10000_log}.

        \subsection{Discussion and Limitations}
            From \Cref{fig:PDF_max_everyN0,fig:PDF_max_everyN0_log}, it can be seen that as $N_0$ increases, the peak of the PDF moves to the right and the range becomes narrower. As $N_0$ increases, the number of samples for searching the maximum value per trial increases. Therefore, a larger value is counted as the maximum value per trial, and the peak moves to the right. The fact that the shape of the PDF changes with $N_0$ suggests that we need to pay attention to the number of samples $N_0$ per trial when estimating the maximum value. 
            %
            \par
            Also, looking at the \Cref{fig:PDF_max_N0_100,fig:PDF_max_N0_100_log}, 
            in terms of the position of the peak, the result obtained by the moment method is in better agreement with the result of MCS than that by Roberts' stochastic averaging method.
            This may be due to the accuracy of the estimation of the tail of the PDF of the roll amplitude by the theoretical calculation, as mentioned in \Cref{sec:exp_discuss}. Therefore, it can be concluded that the moment method provides a better qualitative result for the PDF of the maximum roll amplitude. On the other hand, under conditions where $N_0$ is large, the deviation between the theoretical value and the MCS result becomes large, and the estimation accuracy remains somewhat problematic. In the future, it may be necessary to improve the theoretical estimation method of the PDF of the roll amplitude.
            %
            %

\section{Conclusion}
    %
    The accuracy of the PDF estimation of the parametric roll amplitude in long-crested  irregular waves was validated by comparing the results of each theoretical method with the experimental results. In this study, the experimental results were generally smaller, especially after $20~\mathrm{deg}$, and differed from the results of each theoretical method. In the tail section of the PDF, the PDF result of the moment method is smaller than those obtained by Roberts' stochastic averaging method and ESAM. Therefore, in the considered cases, the moment method is the most suitable for extreme value estimation. The experimental PDF of the roll amplitude had a shape with humps and hollows. We believe that this is due to the change in the yaw angle in the experiment. Further study is needed to develop a model which introduces an additional yaw disturbance and also to improve the experimental method. 
    \par
    Then, the PDFs of the maximum roll amplitude were calculated using the moment method and Roberts' stochastic averaging method, and the results were compared with that using the MCS. The results showed that the moment method gave a better agreement with the MCS results than Roberts' stochastic averaging method, and that the maximum value could be estimated quantitatively. However, it was also confirmed that the deviation from the MCS results becomes larger when $N_0$ is large. We conclude that more accurate estimation of the PDF of the parametric roll amplitude is needed to resolve this issue.
    %
    %
    %

    %
    
\begin{acknowledgements}
    This study was supported by a Grant-in-Aid for Scientific Research from the Japan Society for the Promotion of Science (JSPS KAKENHI Grant \#22H01701).
\end{acknowledgements}

%

\bibliographystyle{spphys}       

\bibliography{main.bib}   

\appendix
\setcounter{equation}{0}%
\setcounter{section}{0}%
\setcounter{subsection}{0}%
\setcounter{figure}{0}%
\setcounter{table}{0}%
\def\theequation{A.\arabic{equation}}
\def\thesection{}
\def\thesubsection{A.\arabic{subsection}}
\def\thefigure{A.\arabic{figure}}
\def\thetable{A.\arabic{table}}
\normalfont
\section{Appendix A. Method of the estimation of GM variation}
\label{sec:appendix}
\subsection{Procedure for calculating GM variation}\label{subsec:method:gm} 
        The following procedure is used to calculate the GM variation ($\Delta\mathrm{GM}$) in irregular waves.
        \begin{enumerate}
            \item Generation of time series for Grim's effective wave\label{item:Grim_t_series}
            \item Equation relating wave elevation and GM Variation \label{item:wave_and_restoring}
            \item Generation of time series of GM variation\label{item:GM_t_series}
        \end{enumerate}
        Each calculation method is explained following.
    
    \subsubsection{Grim's effective wave concept}\label{method_of_Grim}
        
\begin{figure*}[tb]
            \centering
            \includegraphics[width=1\linewidth]{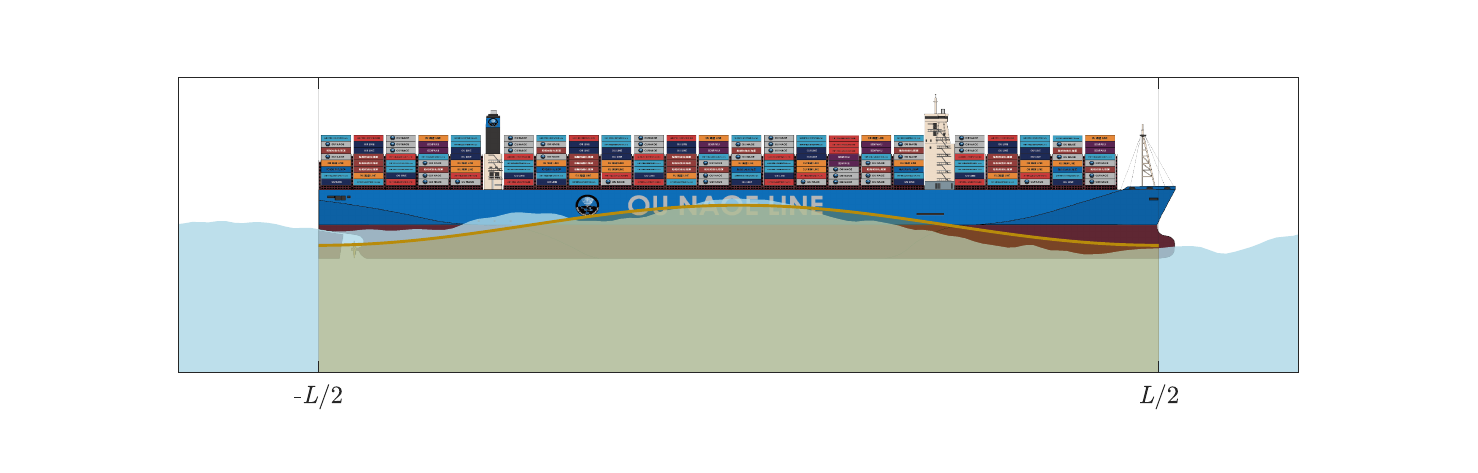}
            \caption{The schematic view of Grim's effective wave}
            \label{fig:Grim figure}
        \end{figure*}

        This is the calculation method performed in \Cref{item:GM_t_series}. Grim's effective wave is the replacement of a spatially irregular waveform around a ship by a single regular wave using the least-squares method, and this regular wave is called the effective wave\cite{Grim1961}. As in \Cref{fig:Grim figure}, this effective wave is assumed to have a wavelength to ship length ratio of 1, and the crest or trough of the wave is in the center of the hull. The amplitude of this effective wave has a linear relationship with ocean wave displacement, while the effect of the wave on the righting lever has a nonlinear and non-memory relationship. As a result, there are no major obstacles to the statistical treatment of the GM variation in irregular waves. 
        The time series of the effective wave displacement $\zeta_{\mathrm{Grim}}(t)$ is calculated by using the ocean wave spectrum $S(\omega)$. 
        Two methods for generating the time series of effective wave is presented by Maruyama et al~\cite{Maruyama2022_moment}. One method is based on the superposition of component waves [Method 1]~\cite{Grim1961}, and the other is based on the application of a linear filter [Method 2]~\cite{Maruyama2022_moment}.
        In this study, in \Cref{sec:result_exp} and \Cref{sec:PDF_maximum}, Method 1 was used on Roberts' stochastic averaging method and ESAM, and Method 2 was used on MCS and the moment method.
        In \Cref{sec:gm_variation}, Method 1 was used.


        \subsubsection{Method of non-memory transformation}\label{method_of_non_memory}
            
\begin{figure}[tb]
                \centering
                \includegraphics[width=0.8\linewidth]{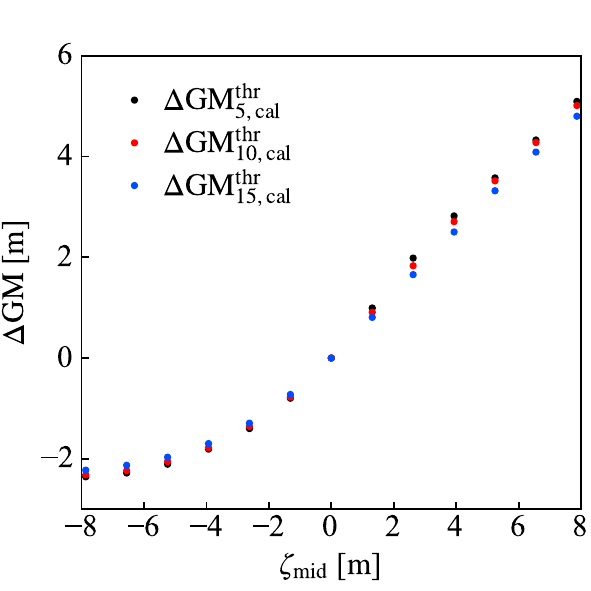}
                \caption{The relationship between $\Delta \mathrm{GM}$ and $\zeta_{\mathrm{mid}}$}
                \label{fig:wavevsGM}
            \end{figure}

\begin{figure}[tb]
                \centering
                \includegraphics[width=0.8\linewidth]{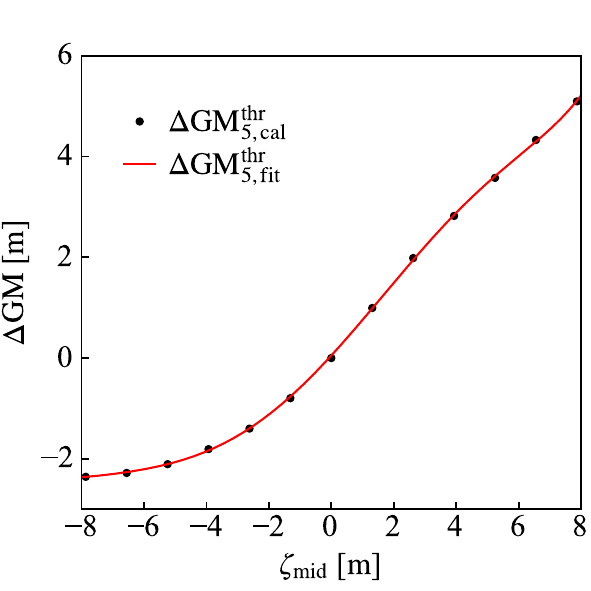}
                \caption{The polynomial approximation of the relationship between $\Delta \mathrm{GM}$ and $\zeta_{\mathrm{mid}}$}
                \label{fig:fig_non_memory}
            \end{figure}

            This is the calculation method performed by \cref{item:wave_and_restoring}. Considering only the wave component based on the Froude-Krylov assumption under quasi-statically balancing heave and pitch in waves, we obtain the equation relating wave displacement and GM variation~\cite{Maruyama2022} in regular waves with the ratio of the wavelength to the ship length of 1. The wave peaks and troughs are assumed to always exist in the center of the hull.
            Using the numerical simulation, we calculate the GM variation $\Delta\mathrm{GM^{thr}_{\phi,cal}}$ when the wave peaks and troughs are located in the center of the hull and the heel angle of the hull is $\phi~[\mathrm{deg}]$. In doing so, calculations are performed for several wave amplitudes $\zeta_{\mathrm{mid}}$. As a result, the relationship diagram in \Cref{fig:wavevsGM} is obtained. Where the wave displacement is positive when the wave trough is located in the center of the hull and it is negative when the wave crest is located in the center of the hull. 
            The original data is plotted by the black dotted line in \Cref{fig:wavevsGM}. The red dotted line in \Cref{fig:fig_non_memory} is a polynomial approximation of the original data as in \Cref{eq:non-memory}, where $N_{\mathrm{p}} = 6$. Using this relation given by \Cref{eq:non-memory} between the wave displacement at the center of the hull and the GM variation, the time series of the effective wave displacement calculated in \cref{item:Grim_t_series} is transformed into the time series of the GM variation.
    
            \begin{equation}
                \Delta \mathrm{GM^{thr}_{\phi,fit}}(\zeta_{\mathrm{mid}}) = \sum_{\mathrm{k}=0}^{N_\mathrm{p}}\mathrm{C}_\mathrm{k}\zeta_{\mathrm{mid}}^{\mathrm{k}} \label{eq:non-memory}
            \end{equation}

\end{document}